\documentclass[letter]{ptptex}

\title{
Distribution of the Riemann Zeros Represented by the Fermi Gas%
}

\author{
Shigenori \textsc{TANAKA}%
}

\inst{
Graduate School of System Informatics, Department of Computational Science, Kobe University, 1-1 Rokkodai, Nada, Kobe 657-8501, Japan
}

\abst{
The multiparticle density matrices for degenerate, ideal Fermi gas system in any dimension are calculated. 
The results are expressed as a determinant form, in which a correlation kernel plays a vital role.
Interestingly, the correlation structure of one-dimensional Fermi gas system is essentially equivalent to that observed for the eigenvalue distribution of random unitary matrices, 
and thus to that conjectured for the distribution of the non-trivial zeros of the Riemann zeta function.
Implications of the present findings are discussed briefly.
}

\PTPindex{062, 013}  

\begin{document}

\maketitle


\vspace{0.5cm}

In this study we consider the noninteracting $N$-particle fermion system at zero temperature in any spatial dimension $d$.
In the following analysis the spin-polarized case is considered, and the volume of the system is taken to be unity for notational simplicity.
The $N$-body density matrix \cite{Mahan} for this Fermi system is then expressed in terms of the product of the Slater determinants 
with the plane waves \cite{Mahan,Wigner} as 

\begin{eqnarray}
\lefteqn{
\rho^{(N)}\left(\mbox{\boldmath $r$}_{1},\mbox{\boldmath $r$}_{2},\cdots,\mbox{\boldmath $r$}_{N}\right)
} \nonumber \\
& & = \frac{1}{N!}\sum_{P^{(N)}}\left[{\rm sgn}P^{(N)}\right]
\exp\left(i\mbox{\boldmath $k$}_{1}\cdot\mbox{\boldmath $r$}_{p_{1}}+
i\mbox{\boldmath $k$}_{2}\cdot\mbox{\boldmath $r$}_{p_{2}}+
\cdots + i\mbox{\boldmath $k$}_{N}\cdot\mbox{\boldmath $r$}_{p_{N}}\right)
\nonumber \\
& & \times \sum_{P'^{(N)}}\left[{\rm sgn}P'^{(N)}\right]
\exp\left(-i\mbox{\boldmath $k$}_{1}\cdot\mbox{\boldmath $r$}_{p_{1}'}-
i\mbox{\boldmath $k$}_{2}\cdot\mbox{\boldmath $r$}_{p_{2}'}-
\cdots - i\mbox{\boldmath $k$}_{N}\cdot\mbox{\boldmath $r$}_{p_{N}'}\right)
\nonumber \\
& & = \frac{1}{N!}\sum_{P^{(N)}}\left[{\rm sgn}P^{(N)}\right]
\exp\left(i\mbox{\boldmath $k$}_{p_{1}}\cdot\mbox{\boldmath $r$}_{1}+
i\mbox{\boldmath $k$}_{p_{2}}\cdot\mbox{\boldmath $r$}_{2}+
\cdots + i\mbox{\boldmath $k$}_{p_{N}}\cdot\mbox{\boldmath $r$}_{N}\right)
\nonumber \\
& & \times \sum_{P'^{(N)}}\left[{\rm sgn}P'^{(N)}\right]
\exp\left(-i\mbox{\boldmath $k$}_{p_{1}'}\cdot\mbox{\boldmath $r$}_{1}-
i\mbox{\boldmath $k$}_{p_{2}'}\cdot\mbox{\boldmath $r$}_{2}-
\cdots - i\mbox{\boldmath $k$}_{p_{N}'}\cdot\mbox{\boldmath $r$}_{N}\right),
\end{eqnarray}

\noindent
where $\mbox{\boldmath $r$}_{i}$ and $\mbox{\boldmath $k$}_{i}$ refer to the spatial coordinates and the wavenumber vectors, respectively;

\begin{equation}
P^{(N)} = \left(
\begin{array}{cccc}
1 & 2 & \cdots & N \\
p_{1} & p_{2} & \cdots & p_{N} 
\end{array}
\right)
\end{equation}

\noindent
and another $P'^{(N)}$ mean the permutations of order $N$, and 

\begin{equation}
{\rm sgn}P^{(N)} = \left\{
\begin{array}{rl}
+1, &\quad\mbox{for even permutation} \\
-1, &\quad\mbox{for odd permutation}
\end{array}\right.
\end{equation}

\noindent
denotes their signature.
The density matrix given by Eq. (1) satisfies the normalization condition as 

\begin{equation}
\int d\mbox{\boldmath $r$}_{1}\cdots d\mbox{\boldmath $r$}_{N} 
\rho^{(N)}\left(\mbox{\boldmath $r$}_{1},\mbox{\boldmath $r$}_{2},\cdots,\mbox{\boldmath $r$}_{N}\right) 
= \frac{1}{N!}\sum_{P^{(N)}}\left[{\rm sgn}P^{(N)}\right]^{2}
= 1.
\end{equation}

\par

Here, noting the identities for $N \times N$ matrices $\mbox{\boldmath $A$}$, its transposition $\mbox{\boldmath $A$}^{T}$, and $\mbox{\boldmath $B$}$ as 

\begin{equation}
\det\mbox{\boldmath $A$}^{T}=\det\mbox{\boldmath $A$}
\end{equation}

\noindent
and

\begin{equation}
\det\mbox{\boldmath $A$}\det\mbox{\boldmath $B$}=\det\mbox{\boldmath $A$}\mbox{\boldmath $B$},
\end{equation}

\noindent
we find

\begin{eqnarray}
\rho^{(N)}\left(\mbox{\boldmath $r$}_{1},\mbox{\boldmath $r$}_{2},\cdots,\mbox{\boldmath $r$}_{N}\right)
&=& \frac{N^N}{N!}\sum_{P^{(N)}}\left[{\rm sgn}P^{(N)}\right]K_{1p_{1}}K_{2p_{2}}\cdots\times K_{Np_{N}}
\nonumber \\
&=& \frac{N^N}{N!}\det\mbox{\boldmath $K$}^{(N)},
\end{eqnarray}

\noindent
where

\begin{equation}
\mbox{\boldmath $K$}^{(N)} = \left(K_{ij}\right)
\end{equation}

\noindent
is an $N \times N$ matrix whose components are 

\begin{equation}
K_{ij} = \frac{1}{N}\sum_{k=1}^{N}\exp\left[i\mbox{\boldmath $k$}_{k}\cdot\left(\mbox{\boldmath $r$}_{i}-\mbox{\boldmath $r$}_{j}\right)\right]
\end{equation}

\noindent
for $1 \le i,j \le N$.
The $n$-body ($1 \le n \le N$) density matrix \cite{Mahan} is then given by

\begin{eqnarray}
\rho^{(n)}\left(\mbox{\boldmath $r$}_{1},\mbox{\boldmath $r$}_{2},\cdots,\mbox{\boldmath $r$}_{n}\right)
&=& \int d\mbox{\boldmath $r$}_{n+1}\cdots d\mbox{\boldmath $r$}_{N}
\rho^{(N)}\left(\mbox{\boldmath $r$}_{1},\mbox{\boldmath $r$}_{2},\cdots,\mbox{\boldmath $r$}_{N}\right)
\nonumber \\
&=& \frac{(N-n)!}{N!}N^{n}\sum_{P^{(n)}}\left[{\rm sgn}P^{(n)}\right]K_{1p_{1}}K_{2p_{2}}\cdots\times K_{np_{n}}
\nonumber \\
&=& \frac{(N-n)!}{N!}N^{n}\det\mbox{\boldmath $K$}^{(n)}
\end{eqnarray}

\noindent
with an $n \times n$ matrix $\mbox{\boldmath $K$}^{(n)}$.
It is noted that the prefactor before $\det\mbox{\boldmath $K$}^{(n)}$ in Eq. (10) becomes unity for finite $n$ and $N \to \infty$.

\par

A problem in the following is to find an explicit expression for the correlation kernel,

\begin{equation}
K\left(\mbox{\boldmath $r$}\right) =
\frac{1}{N}\sum_{\mbox{\boldmath $k$}}\exp\left(i\mbox{\boldmath $k$}\cdot\mbox{\boldmath $r$}\right)n_{\mbox{\boldmath $k$}},
\end{equation}

\noindent
in any dimension $d$, which represents Eq. (9) in the limit of infinite $N$.
Here, the $\mbox{\boldmath $k$}$ summation is expressed by

\begin{equation}
\sum_{\mbox{\boldmath $k$}} = \frac{1}{(2\pi)^{d}}\int d\mbox{\boldmath $k$},
\end{equation}

\noindent
and for the degenerate Fermi gas, 

\begin{equation}
n_{\mbox{\boldmath $k$}} = \theta\left(k_{\rm{F}}-\vert\mbox{\boldmath $k$}\vert\right)
\end{equation}

\noindent
refers to the occupation number density in the wavenumber space 
with $\theta(x)$ and $k_{\rm{F}}$ being the step function and the Fermi wavenumber, respectively.

\par 

The occupation number density satisfies the normalization condition, 

\begin{equation}
\sum_{\mbox{\boldmath $k$}}n_{\mbox{\boldmath $k$}} = N.
\end{equation}

\noindent
Therefore, the Fermi wavenumber $k_{\rm{F}}$ is given by 

\begin{equation}
\frac{1}{(2\pi)^{d}}V_{d}\left(k_{\rm{F}}\right) = N,
\end{equation}

\noindent
where

\begin{equation}
V_{d}\left(k_{\rm{F}}\right) = C_{d}k_{\rm{F}}^{d}
\end{equation}

\noindent
with

\begin{equation}
C_{d} = \frac{\pi^{d/2}}{\Gamma\left(\frac{d}{2}+1\right)}
\end{equation}

\noindent
is the volume of the $d$-dimensional sphere with the radius of $k_{\rm{F}}$;
$\Gamma(s)$ is the gamma function \cite{Abra,Grad}.
The Fermi wavenumber is thus expressed as 

\begin{equation}
k_{\rm{F}} = 2\sqrt{\pi}\left[\Gamma\left(\frac{d}{2}+1\right)N\right]^{1/d}.
\end{equation}

\par

By choosing the direction of $\mbox{\boldmath $r$}$ in parallel with the $d$-th component of the wavenumber vector $\mbox{\boldmath $k$}$, 
Eq. (11) is expressed as 

\begin{eqnarray}
K\left(\mbox{\boldmath $r$}\right) &=&
\frac{1}{N(2\pi)^{d}}\int\nolimits_{-1}^{1}dt\ k_{\rm{F}}\exp\left(ik_{\rm{F}}rt\right)V_{d-1}\left(k_{\rm{F}}\sqrt{1-t^2}\right)
\nonumber \\
&=& \frac{C_{d-1}k_{\rm{F}}^{d}}{N(2\pi)^d}
\int\nolimits_{-1}^{1}dt\exp\left(ik_{\rm{F}}rt\right)\left(1-t^2\right)^{\frac{d-1}{2}}
\end{eqnarray}

\noindent
with $r = \vert \mbox{\boldmath $r$}\vert$.
Then, employing Poisson's formula \cite{Abra,Grad}, 

\begin{equation}
\int\nolimits_{-1}^{1}dt\exp(izt)\left(1-t^2\right)^{\nu-\frac{1}{2}} 
= \frac{\sqrt{\pi}\Gamma\left(\nu+\frac{1}{2}\right)}{(z/2)^{\nu}}J_{\nu}(z),
\end{equation}

\noindent
we find an explicit expression for the correlation kernel as

\begin{equation}
K\left(\mbox{\boldmath $r$};\nu\right) = \Gamma(\nu + 1)\left(\frac{2}{k_{\rm{F}}r}\right)^{\nu}J_{\nu}\left(k_{\rm{F}}r\right)
\end{equation}

\noindent
with $\nu =d/2$, where $J_{\nu}(z)$ is the Bessel function of the first kind of order $\nu$ \cite{Abra,Grad}.
It is remarked that, though Eq. (21) has been derived for integral values of spatial dimension $d$, 
$K(\mbox{\boldmath $r$};\nu)$ may be regarded as a continuous function of the auxiliary order variable $\nu$, 
which follows from its series representation \cite{Abra,Grad}.

\par

Let us here consider the case of $d = 3$ and $\nu = 3/2$.
The correlation kernel then reads 

\begin{eqnarray}
K\left(\mbox{\boldmath $r$}\right) &=& 
\Gamma\left(\frac{5}{2}\right)\left(\frac{2}{k_{\rm{F}}r}\right)^{3/2}J_{3/2}\left(k_{\rm{F}}r\right)
\nonumber \\
&=& \frac{3}{k_{\rm{F}}r}j_{1}\left(k_{\rm{F}}r\right) 
\nonumber \\
&=& \frac{3}{\left(k_{\rm{F}}r\right)^{3}}\left(\sin k_{\rm{F}}r - k_{\rm{F}}r\cos k_{\rm{F}}r\right),
\end{eqnarray}

\noindent
where

\begin{equation}
j_{n}(z) = \left(\frac{\pi}{2z}\right)^{1/2}J_{n+\frac{1}{2}}(z)
\end{equation}

\noindent
is the spherical Bessel function of the first kind of order $n$ \cite{Abra,Grad}.
This expression, Eq. (22), combined with Eq. (10) in the case of two-body density matrix, reproduces a well-known result \cite{Mahan,Wigner} for the pair distribution function 
of the three-dimensional Fermi gas.

\par

In the case of $d = 1$ and $\nu = 1/2$, we find 

\begin{eqnarray}
K(r) &=& 
\Gamma\left(\frac{3}{2}\right)\left(\frac{2}{k_{\rm{F}}r}\right)^{1/2}J_{1/2}\left(k_{\rm{F}}r\right)
\nonumber \\
&=& j_{0}\left(k_{\rm{F}}r\right) 
\nonumber \\
&=& \frac{\sin k_{\rm{F}}r}{k_{\rm{F}}r}.
\end{eqnarray}

\noindent
Recalling Eq. (10), this correlation structure is essentially the same as that for the eigenvalues of random matrices in the Gaussian 
unitary ensemble \cite{Dyson1,Dyson2,Dyson3}.
Interestingly, it has been known that this type of correlation structure may hold also for the distribution of zeros in the Riemann zeta function \cite{Titch}. 

\par

The Riemann zeta function \cite{Titch} for complex variable $s$ is defined by 

\begin{equation}
\zeta(s) = \sum_{n=1}^{\infty}\frac{1}{n^s} = \prod_{p}\left(1-p^{-s}\right)^{-1}
\end{equation}

\noindent
for Re\ $s > 1$, where $n$ and $p$ mean the natural numbers and the prime numbers, 
respectively. After the analytic continuation over the whole complex plane, the $\zeta(s)$ has non-trivial zeros in the critical strip, 
$0 < $ Re\ $s < 1$, and the 
Riemann hypothesis \cite{Titch} states that all of them lie on the critical line Re\ $s = 1/2$; that is, 

\begin{equation}
\zeta\left(\frac{1}{2}+it\right) = 0
\end{equation}

\noindent
has non-trivial solutions only when $t = t_{j}\ (j=1,2,\cdots)$ are real.

\par

The mean density of the non-trivial zeros of $\zeta(s)$ increases logarithmically with height $t$ up to the critical line. We then define unfolded zeros by 

\begin{equation}
w_{j} = \frac{t_{j}}{2\pi}\log\frac{t_{j}}{2\pi}.
\end{equation}

\noindent
It has then been conjectured \cite{Montgomery,Odlyzko} that the pair correlation function of the unfolded zeros for $j \to \infty$ has a form,  

\begin{equation}
R^{(2)}(w) = 1 - \left(\frac{\sin\pi w}{\pi w}\right)^{2},
\end{equation}

\noindent
which is analogous to that for the one-dimensional ($d = 1$) Fermi gas, that is, Eqs. (24) and (10) for $n=2$. 
This conjecture has also been generalized for all the $n$-point correlations as a determinant form 
analogous to Eq. (10) \cite{Rudnick,Bogomolny1,Bogomolny2,Keating}.

\par

Thus, the case of $\nu = d/2 \to 1/2$ in the Fermi gas system gives a special correlation structure observed in the random unitary matrices 
and the Riemann zeta function.
Therefore, the behaviors of the multiparticle correlations of the Fermi gas system with the correlation kernel $K\left(\mbox{\boldmath $r$};\nu\right)$ 
at and around $\nu = 1/2$ may provide useful information about the random matrices and the zeta function, regarding $K\left(\mbox{\boldmath $r$};\nu\right)$ 
as a continuous function of $\nu$.
Another challenge, which would be more ambitious, is to look for a family of functions whose zero distributions are described  by the correlation functions 
given by Eqs. (10) and (21) for arbitrary values of $\nu$.

\par

In summary, it has been revealed in this study that the correlation structure similar to those observed in the eigenvalues of the random unitary matrices 
and in the Riemann zeros is embedded in the Fermi gas system as well.


%


\begin{thebibliography}{99}
  
%
%
\bibitem{Mahan} G.D. Mahan, {\em Many-Particle Physics} (Plenum Press, New York, 1990).
\bibitem{Wigner} E. Wigner and F. Seitz, Phys. Rev. \textbf{43} (1933), 804.
\bibitem{Abra} M. Abramowitz and I.A. Stegun, {\it Handbook of Mathematical Functions} (Dover Publications, New York, 1965).
\bibitem{Grad} I.S. Gradshteyn and I.M. Ryzhik, {\it Table of Integrals, Series, and Products} (Academic Press, San Diego, 1994).
\bibitem{Dyson1} F.J. Dyson, J. Math. Phys. \textbf{3} (1962), 140.
\bibitem{Dyson2} F.J. Dyson, J. Math. Phys. \textbf{3} (1962), 157.
\bibitem{Dyson3} F.J. Dyson, J. Math. Phys. \textbf{3} (1962), 166.
\bibitem{Titch} E.C. Titchmarsh and D.R. Heath-Brown, {\it The Theory of the Riemann Zeta-function} (Clarendon Press, Oxford, 1986).
\bibitem{Montgomery} H.L. Montgomery, Proc. Symp. Pure Math. \textbf{24} (1973), 181.
\bibitem{Odlyzko} A.M. Odlyzko, in {\it Dynamical, Spectral, and Arithmetic Zeta Functions}, edited by M. van Frankenhuysen and M.L. Lapidus 
(Am. Math. Soc., Contemporary Math. Series, No. 290, 2001), p. 139.
\bibitem{Rudnick} Z. Rudnick and P. Sarnak, Duke Math. J. \textbf{81} (1996), 269.
\bibitem{Bogomolny1} E.B. Bogomolny and J.P. Keating, Nonlinearity \textbf{8} (1995), 1115.
\bibitem{Bogomolny2} E.B. Bogomolny and J.P. Keating, Nonlinearity \textbf{9} (1996), 911.
\bibitem{Keating} J.P. Keating and N.C. Snaith, J. Phys. A \textbf{36} (2003), 2859.


\end{thebibliography}
\end{document}